\documentclass[journal]{vgtc}                

\ifpdf
  \pdfoutput=1\relax                   
  \usepackage{graphicx}                
  \DeclareGraphicsExtensions{.pdf,.png,.jpg,.jpeg} 
\else
  \usepackage{graphicx}                
  \DeclareGraphicsExtensions{.eps}     
\fi%

\graphicspath{{figures/}{pictures/}{images/}{./}} 

\usepackage{microtype}                 
\PassOptionsToPackage{warn}{textcomp}  
\usepackage{textcomp}                  
\usepackage{mathptmx}                  
\usepackage{times}                     
\usepackage{cite}                      
\usepackage{tabu}                      
\usepackage{booktabs}                  
\usepackage{paralist}
\usepackage{enumitem}

\usepackage{multirow}
\usepackage{color}
\usepackage{amsmath}
\usepackage{bm}
\usepackage{xspace}
\usepackage{setspace}
\usepackage{amsfonts}
\usepackage{dblfloatfix}

\onlineid{1229}

\vgtccategory{Research}
\vgtcpapertype{application/design study}

\title{PassVizor: Toward Better Understanding of the Dynamics of Soccer Passes}

\author{Xiao Xie, Jiachen Wang, Hongye Liang, Dazhen Deng, Shoubin Cheng, Hui Zhang, Wei Chen, Yingcai Wu}
\authorfooter{
\item
 X.Xie, J.Wang, H.Liang, D.Deng, W.Chen, and Y.Wu are with the State Key Lab of CAD\&CG, Zhejiang University. E-mail: \{xxie, wangjiachen, lianghongye, dazhendeng, ycwu\}@zju.edu.cn, chenwei@cad.zju.edu.cn. Y.Wu is the corresponding author.
\item
 H.Zhang and S.Cheng are with the Department of Sport Science, Zhejiang University. E-mail: zhang\_hui@zju.edu.cn, shoubincheng@outlook.com.
}

\shortauthortitle{Xiao Xie \MakeLowercase{\textit{et al.}}: PassVizor: Toward Better Understanding of the Dynamics of Soccer Passes}

\abstract{
In soccer, passing is the most frequent interaction between players and plays a significant role in creating scoring chances. Experts are interested in analyzing players’ passing behavior to learn passing tactics, i.e., how players build up an attack with passing. Various approaches have been proposed to facilitate the analysis of passing tactics. However, the dynamic changes of a team’s employed tactics over a match have not been comprehensively investigated. To address the problem, we closely collaborate with domain experts and characterize requirements to analyze the dynamic changes of a team’s passing tactics. To characterize the passing tactic employed for each attack, we propose a topic-based approach that provides a high-level abstraction of complex passing behaviors. Based on the model, we propose a glyph-based design to reveal the multi-variate information of passing tactics within different phases of attacks, including player identity, spatial context, and formation. We further design and develop PassVizor, a visual analytics system, to support the comprehensive analysis of passing dynamics. With the system, users can detect the changing patterns of passing tactics and examine the detailed passing process for evaluating passing tactics. We invite experts to conduct analysis with PassVizor and demonstrate the usability of the system through an expert interview.
}


\CCScatlist{ 
 \CCScat{K.6.1}{Management of Computing and Information Systems}%
{Project and People Management}{Life Cycle};
 \CCScat{K.7.m}{The Computing Profession}{Miscellaneous}{Ethics}
}

\vgtcinsertpkg

\begin{document}



\maketitle

\begin{spacing}{0.97}

\section{Introduction}\label{sec:introduction}

Soccer is a globally popular sport. Considering the broad impact of soccer, experts are eager to improve team performance by collecting and analyzing soccer data. In particular, passing has received considerable attention in soccer analysis. 
The most popular passing analysis is to construct a passing network in which nodes encode players and edges encode the number of passes between players. This kind of analysis can be frequently seen in the match reports from sports analytics websites like Opta\cite{opta} and STATS\cite{stats}. Based on the network, analysts can do different kinds of analysis, such as finding important playmakers by searching for high-degree nodes and detecting frequent passes by finding strong edges. 

Particularly, many analysts desire to learn valuable attacking patterns in terms of passing. We refer attacking patterns as a characterization of how players build up an attack. For example, multiple counter-attacks may contain a same pattern, passing between Midfielder A and left Winger B. Finding this kind of patterns can help analysts understand the main focus of attacks. Moreover, uncovering the dynamics of passing patterns can facilitate the study of coaches' adjustments of strategies. For example, a shift from offensive passing to defensive passing may represent a strategy adjustment for holding the lead. Yet the passing pattern detection can be hardly accomplished with the passing network. Passes in different soccer phases are aggregated into a network, thereby losing the information of building up an attack.

\begin{figure}
    \centering
    \includegraphics[width=1\linewidth]{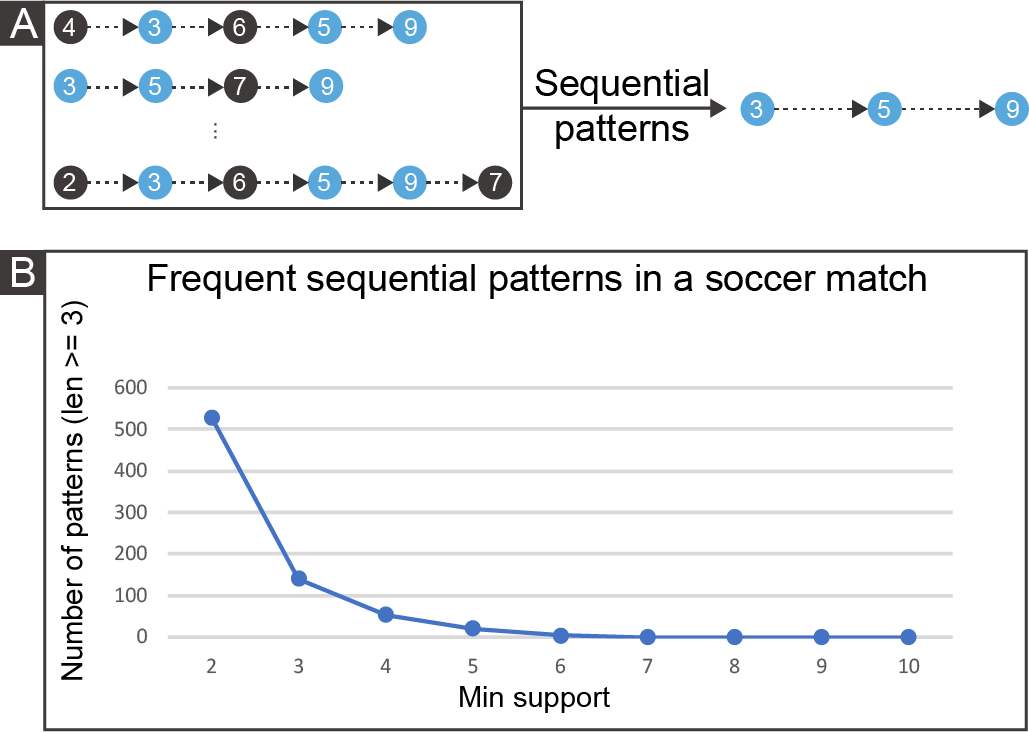}
    \caption{(A) The sequential pattern mining of passing. Passing in a soccer phase is regarded as a sequence of players. The sequential pattern mining of passing is to discover the frequent subseqences. (B) The results of sequential pattern mining on a soccer match. We used the \textit{prefixspan} to detect the sequential patterns. The detected patterns are not informative as most of the pattern lengths are limited to two. The restriction of sequence orders contradicts the dynamic match process.}~\label{sequential}
    \vspace{-8mm}
\end{figure}

In this paper, we aim to help analysts efficiently analyze the passing at a detailed level to support in-depth analysis. We have interviewed with four soccer experts to collect their requirements of passing analysis. Based on the requirements, we propose a novel visual analytics system, PassVizor, to achieve our goal. PassVizor introduces the concept of topics into the detection of passing patterns. Recent works \cite{DecroosHD18} transform passes to sequences of roles (e.g., forwards) and apply sequential pattern mining  (Fig.~\ref{sequential}(A)) to detect frequent subsequences as patterns. However, when we transform the pass to the sequence of actual players rather than player roles (Fig.~\ref{sequential}(B)) for preserving the identity, most of the pattern lengths are limited to two, which is meanless for analysis. To solve this problem, we turn to topic modeling, another effective approach for characterizing sequence data. This is inspired by the similar structure between passing sequences and sentences where the co-occurrence of words follows the constrain of semantics and the co-occurrence of players follows the constrain of strategies. Hence, we use latent topics to encode the latent passing patterns under complex passing sequences.

PassVizor further utilizes a glyph-based design to visualize the multi-variate passing context information.
Existing visualizations have proposed different visual designs to demonstrate passes in a soccer phase. However, a soccer match usually involves hundreds of phases. How to help analysts quickly navigate through multiple soccer phases remains an important issue to be tackled. To solve this problem, we leverage the visual abstractness of glyphs and propose glyph designs for the context information of passing, such as the spatial information and the team formation. Multiple soccer phases can be juxtaposed for analysis.



The combination of passing pattern detections and concise visualizations of passing enables analysts to realize the change of passing patterns and to interpret these kinds of changes. The main contributions of this work are as follows:

\begin{compactitem}
    \item A hierarchical problem characterization that summarizes requirements from match-level to player-level for analyzing the dynamic changes of passing.
    \item A topic-based approach for modeling the passing sequences and detecting passing patterns that can preserve the collaboration information.
    \item An interactive visual analytics system for supporting a comprehensive analysis of the dynamics of passing.
\end{compactitem}
\section{Related Work}

Our work is mainly related to two parts of works, namely, the analysis of passing and the visualization of soccer.
\subsection{Analysis of Passing}
In recent years, soccer analysts have proposed a series of works to investigate players' passing.
Based on the adopted theories, existing methods can be mainly divided into two categories, namely, the \textit{network-based} and the \textit{sequence-based}.

\textit{Network-based.} The network-based methods aggregate passes to create a directed graph in which nodes represent players and edges represent the frequency of passes between players.
With traditional graph analysis methods, experts can extract passing patterns from the derived passing network method \cite{clemente2014using, 1206.6904, clemente2015midfielder}.
For example, to explain why Spain won the World Cup in 2010, Pe{\~n}a and Touchette \cite{1206.6904} constructed passing networks for Spain and Netherlands.
The node centralities showed that the players of Spain, relative to those of the Netherlands, were equally involved in the passing network and thus presented a steady and consolidated playing style. As aggregating passes of different times may obscure important variations, the sliding window method has been applied to create temporal passing networks \cite{clemente2015using, gonccalves2017exploring, yamamoto2011common, cotta2013network, duarte2013capturing}.
For example, Cotta et al. \cite{cotta2013network} computed several network measures for passing networks at different time steps to inspect the evolution.
Yamamoto et al. \cite{yamamoto2011common} visualized the numbers of triangle structures in each 5-min timespan of a soccer match to show the dynamics.
However, as a result of the aggregation process, the network-based method break a series of passes into multiple independent passes between players.
This characteristic poses challenges for analysts when investigating multi-step passing patterns (e.g., wall passing).
The sequence-based methods were therefore introduced to preserve these details.

\textit{Sequence-based.} The sequence-based methods regard a series of consecutive passes in a soccer phase as a sequence where each element represents for an action of players (e.g., passing) \cite{1506.07768}.
With the sequence representation of passing, Gyarmati et al. \cite{1409.0308} extracted three-step passing patterns of different teams and found that FC Barcelona tended to pass the ball back and forth, which is a playing style that differs from those used by others.
Lucey et al. \cite{LuceyOCRM13} proposed a technique called occupancy maps to characterize the passing patterns in different regions of the field.
By comparing the frequent sequence patterns that lead to goals, Bekkers and Dabadghao \cite{bekkers2017flow} provided a list of case studies, such as the effect of trading players and the replacement of coaches.
On the basis of the ball-movement data, several works \cite{DecroosHD18} further detected frequent trajectory patterns of passing by using techniques such as dynamic time warping.
These methods have been widely used to extract typical passing patterns to characterize players' passing behaviors.
However, this type of analysis largely discards the temporal dynamics of passing (e.g., when a team changes their passing behaviors) as well as other important context information (e.g., defense of the opponent).
Therefore, we develop PassVizor to help experts discover and understand such dynamics of passing behaviors.

\subsection{Visualization of Soccer}
Researchers have proposed a set of visualization techniques and tools to analyze different sports data, such as the basketball \cite{Chen2016nba, Cervone2014, Losada2016}, the baseball\cite{DBLP:journals/cgf/OnoDS18}, and racquet sports \cite{ittvis, wu2020visual, ye2020shuttlespace, tacsimur}.
Specifically, a number of approaches have been introduced to present and analyze soccer data from different aspects.
For situation awareness, Legg et al. \cite{LeggCPJLGC12} created glyphs of different sports events and visualized ongoing matches in real-time.
Soccer Scoop \cite{Rusu2011iv, Rusu2010iv} developed a visualization tool for comparing performances of players.
Sacha et al. \cite{Sacha2017cgf} provided a visual abstraction method to address the issue of visual clutter in visualizing the massive trajectories of players, which eased the detection of movement patterns.
Andrienko et al. \cite{Andrienko2017pressure} concentrated on evaluating the defense of players, contributed an approach to compute the pressure exerted by defenders, and applied heat map-based visualizations to present the values.
In order to strengthen the traditional video analysis, Stein et al. \cite{Stein2017tvcg} designed an automatic method that can embed the visualizations of players' regions and movements into videos.
In addition to advanced techniques, several visual analytics systems have been proposed for in-depth analysis.
SoccerStories \cite{soccerstories} divided a match into different phases based on players' actions and designed an interface to support the navigation of phases.
Tailored facet views were provided to investigate the details of different actions (e.g., shooting and passing).
Janetzko et al.\cite{Janetzko2014vast, Stein2015ijgi} visualized a time series of multiple soccer features that capture the complex soccer context, which facilitated the player-centric performance analysis.  
Wu et al. \cite{forvizor} contributed a Spatio-temporal design to visualize formation changes in a match and developed a system called ForVizor to support comprehensive analysis.

Despite the diverse focuses of soccer visualizations, players' passing behaviors were presented in most previous works as a necessary part of soccer analysis.
For example, SoccerStories \cite{soccerstories} proposed multiple designs, e.g., adjacency matrix, node-link diagram, and hive plot, to visualize the passing behavior in a soccer phase.
Multiple soccer phases that involve shooting events were juxtaposed for the investigation of a whole match.
However, according to experts, visualizing these phases is not sufficient as most phases are disrupted by the defense and only a few phases end with shootings.
These disrupted phases also involve valuable information on players' passing behaviors and need to be analyzed concurrently.
These drawbacks pose challenges to existing visualizations because juxtaposing hundreds of phases could lead to a heavy cognitive load for users. Moreover, how to visually connect these soccer phases to show the dynamics of passing behaviors remains unsolved. Therefore, we propose PassVizor, an interactive system to support an in-depth analysis of evolving passing behaviors.

\begin{figure*}
	\centering
	\includegraphics[width=1\linewidth]{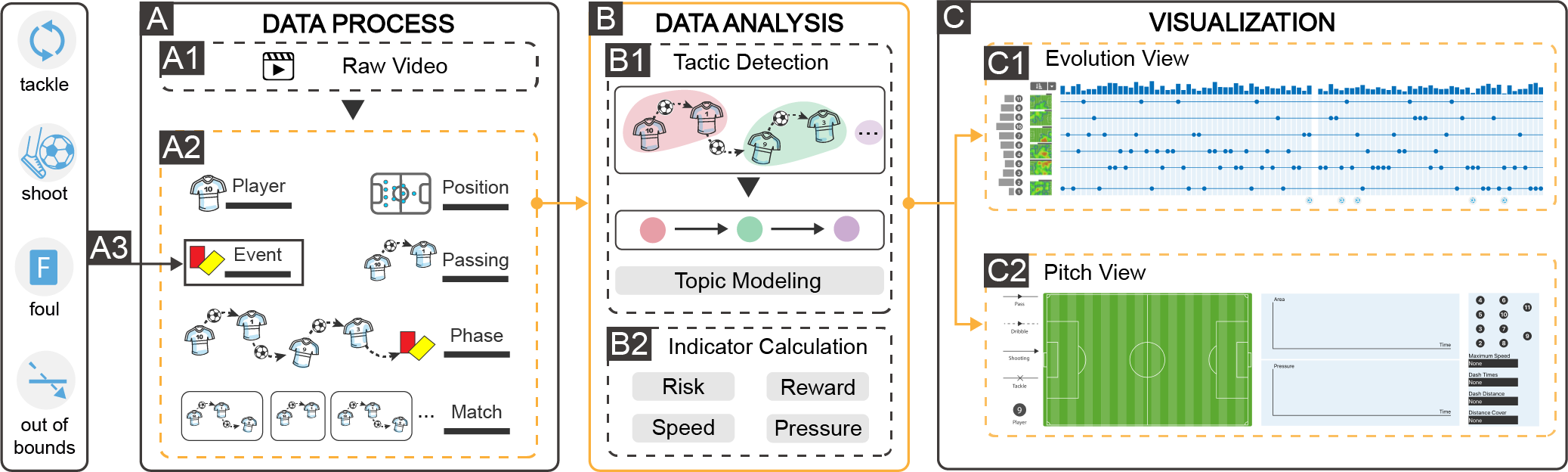}
		\caption{The pipeline of PassVizor. (A) Data processing includes (A1) obtaining the position data from videos and (A2) manually labeling the passing event and dividing the soccer match into multiple soccer phases. (B) Data analysis includes (B1) detecting passing patterns and (B2) computing statistical indicators. (C) Visualization system includes (C1) an evolution view for showing the dynamic change of passing behaviors and (C2) a pitch view for showing the detailed passing in each soccer phase.}~\label{pipeline}
		\vspace{-8mm}
\end{figure*}

\section{Background}
In this section, we first describe the interview process with domain experts.
We then provide a problem characterization of passing analysis based on the interview.

\subsection{Interviews}
We collaborated with four soccer experts for one year to develop a visual analytics system for analyzing the dynamics of soccer passing. 
The four experts included one soccer coach (E1, with a coaching certificate issued by a continental football confederation), one senior sports analyst (E2), and two PhDs majoring in sports science (E3 and E4, both are former professional players of a top soccer league in a country).
To understand domain problems, we collected related works and interviewed experts through weekly meetings.
During the meetings, experts stated that traditional video analysis approaches are cumbersome to use when investigating passing behaviors. E1 and E2 emphasized that revealing the dynamic changes of passing behaviors can facilitate the understanding of the adjustment of a team. The outcomes can help answer questions such as, \textit{When would a team change to a defensive passing?} and \textit{What kind of defensive passing would they employ?} Moreover, E3 and E4 shared their experiences in the decision-making process related to passing. 

\subsection{Results}
\label{requirement}
Inspired by experts' comments and feedback, we identified a set of important questions and problems. We further distilled the problems into the following requirements.

\begin{itemize}
\item \textbf{Match-level Summarization.}

    \begin{compactenum}[M1]
        \item \textit{How many kinds of passing patterns do a team involve?} As a team could have hundreds of phases in a match, a clear overview of passing patterns can be a good starting point for experts to conduct an in-depth analysis of passing. For example, they can quickly learn which kinds of passing patterns have frequently occurred.

        \item \textit{What are the characteristics of each passing pattern?} The demonstration of passing patterns' characteristics is also required. This can help experts quickly focus on their interesting passing patterns. Specifically, revealing characteristics such as involved players, active spatial regions, and frequent passes can facilitate the understanding of the tactical intentions of passing.
    \end{compactenum}

\item \textbf{Phase-level Analysis.}

    \begin{compactenum}[P1]
        \item \textit{How do passing behaviors evolve over phases?} A team could dynamically change the passing behaviors (e.g., from defensive possession to aggressive attacking) according to the game situation and performance. Therefore, tracing the evolution of passing behaviors empowers experts to realize the decision making process of coaches and players.
        \item \textit{How do passing behaviors change according to defenses?} Experts state that players' passing behaviors are related to defenses. Generally, defenders may apply different defensive strategies (e.g., man-marking) in a match. Experts are eager to know how a team copes with different defenses through changing passing behaviors.
        
        \item \textit{How do passing behaviors benefit the team?} Different passing behaviors could lead to different results for a team (e.g., offside, corner, and shooting). Experts want to learn the efficiency of players' passing and figure out what kind of passing can create better opportunities.
    \end{compactenum}

\item \textbf{Individual-level Investigation.}

    \begin{compactenum}[{I}1]
        \item \textit{How does a player complete a pass?} Passes can have different appearances. Players can dribble the ball before passing or they can complete the pass by a one-touch play. The different processes of passing can cause different effects on the attack. Hence, inspecting such processes is essential for justifying and evaluating the passing behavior. 

        \item \textit{What are the characteristics of each player?}  The different passing skills and personalities of players result in different types of passes. Thus, experts are eager to investigate the characteristics of each player, so as to learn players' strengths, weakpoints, and compatibility with team strategies. This knowledge can further ease tasks such as the lineup selection and player substitutions.

        
    \end{compactenum}
\end{itemize}

\subsection{Data Collection}
We prepare our data according to the requirements. There are two types of data that we need for conducting passing analysis, namely, the position data and the event data (e.g., passes, goals, and shoots). We use a semi-automatic method to collect these data from raw videos. For the position data, the methodology of data collection is similar to ForVizor \cite{forvizor} and requires nearly 6 hours for collecting all the position data of a soccer game by two users. Specifically, we have an interface that allows users to inspect the object tracking result and interactively adjust the tracking result when the tracking target is missed or mistakenly labeled. For the event data, it requires nearly 90 minutes for collecting the soccer event data of a soccer game by one user. Users need to watch the video and stop when an event occurs. They further fill the information of each event with the interface. Users do not need specific expertise for collecting the data.

\section{System Overview}
We designed PassVizor, a visual analytics system, to help analysts conduct passing analysis.
The system supports users in learning the dynamic-changing pattern of passing behaviors, detecting valuable passes, and understanding the decision-making process of each pass. The system comprises of two components, namely, a pattern detection component (implemented by Python) (Fig.~\ref{pipeline} (B)) and a visualization component (based on Vue.js 2.0 framework) (Fig.~\ref{pipeline} (C)). Specifically, the pattern detection component utilizes a topic-based approach (Fig.~\ref{pipeline} (B1)) to detect passing patterns. The visualization component (Fig.~\ref{pipeline} (C)) is further developed to support the analysis of passing behaviors.


\section{Passing Modeling}

In this section, we introduce an efficient topic-based model to detect passing patterns.
We first provide an introduction to the data structure.
We then show an overview of the passing pattern detection and explain why we decide to use topic modeling. Finally, we demonstrate how we apply a topic modeling method to detect the passing patterns.

\label{data}
A soccer match is composed of two teams, each with 11 players.
In this work, we focus on analyzing the dynamics of passing behaviors in a match.
To obtain the passing data, we first collected players' positions per frame (Fig.~\ref{pipeline} (A)) from the raw video.
We then manually labeled the passing (Fig.~\ref{pipeline} (A2)) and the subsequent events of passing (Fig.~\ref{pipeline}(A3)).
We further derived a sequence of soccer phases, in which a team conducts an attack by a series of consecutive passes.
The data is structured as follows.

\begin{compactitem}
    \item \textbf{Player} is denoted as $v_i, \ i\in\{1, \dots ,11\}$. 
    \item \textbf{Pass} is denoted by a tuple $e = (v_p, v_r, t_p, t_r)$, where $v_p$ denotes the player that passes the ball and $v_r$ denotes the player that receives the ball. $t_p$ and $t_r$ denote the timestamp of passing and receiving the ball, respectively. The passes in a game are arranged by time as $\{e_1,e_2,\cdots,e_n\}$, where $n$ is the number of passes in the game. 
    
    \item \textbf{Phase} is defined as consecutive passes without being interrupted. For a team, a phase begins when they gain control of the ball and ends when they lose control of the ball. We follow the phase definition of previous works \cite{decroos2018automatic}. We denote each phase as $s_i = (e_j, e_{j+1}, \dots , e_{j+k})$, indicating that the phase consists of the passes from the $j^{th}$ to the $(j+k)^{th}$ ($j,k > 0$ and $j+k\leq n$). The rule of deriving soccer phases can be found in Fig.~\ref{pipeline} (A).
    \item \textbf{Match} is composed by a list of soccer phases $M = (s_1, s_2, \dots , s_m)$, where $m$ is the amount of the phases.
\end{compactitem}

\subsection{Passing Pattern Detection}
\label{feature}

Researchers have proposed different methods to accomplish the detection task.
In recent years, a popular method is to transform the detection of passing patterns to a task of mining frequent sequential patterns. 
The sequential pattern is defined as a subsequence $(a_{i_1}, \dots, a_{i_k})$ where $1 \leq i_1 < i_2 < \dots < i_k \leq n$. Hence, a frequent sequential pattern is a sequential pattern that commonly occurs with a frequency higher than a threshold (Fig.~\ref{sequential}).
With this definition, Gyarmati et al. \cite{1409.0308} defined passing patterns as a sequential pattern of player roles (i.e., $a_i \in \{midfielder, forward, backward\}$) and tried to investigate the passing relation between different player roles. Decroos et al. \cite{DecroosHD18} focused on the spatial positions of passing and defined $a_i$ as a tuple of events and its spatial regions. 
Despite the usefulness, these methods encounter problems when applied to a detailed-level analysis of passing since the personal information of players is neglected (Fig.~\ref{sequential}). This poses difficulties for experts to communicate the finding from passing analysis with existing studies such as the profiling of players.

Therefore, we aim to provide a new passing pattern detection method that can address this limitation. We have surveyed the works of pattern mining in sequence data and find that the task of passing pattern detection is similar to the topic modeling of text sequences.
Specifically, the \textit{co-occurrence} of players, which is what we can observe in the passing data, is the result of coaches' strategies (which is latent in the data).
This is similar to topic modeling, which uses the observed co-occurrence of words to encode the concept of latent topics. 

During the exploration of passing modeling, we presented this co-occurrence based passing pattern to our experts.
Experts commented that this passing pattern definition is consistent with their experiences.
Following coaches' strategies, players usually have a tendency to pass. 
For example, when applying the long passing strategy, players of guards would tend to pass the ball to the center-forward directly to conduct an attack. Such attacks would cause a co-occurrence of the forwards and the guards, while the midfielders would be seldom seen in those attacks. Hence, different strategies would lead to different co-occurrence patterns and finding these patterns can lead to a deeper understanding of the strategies. Therefore, we decide to use topic modeling to detect passing patterns.

We use $p_T(v_i)$ to represent the probability that the player $v_i$ is involved in the pattern $T$. Based on the player co-occurrence, a passing pattern is defined as a tuple of each player's participation probability,
\begin{flalign}
  T = (p_T(v_1),p_T(v_2),\dots,p_T(v_{11})),
\end{flalign}
where $0\leq p_T(v_{i}) \leq 1$.
With this definition, the order information is ignored. Nevertheless, this can improve the robustness of passing patterns since we can hardly see a strictly consistent order of players in multiple passing sequences. 

\begin{figure}[htb]
	\centering
	\includegraphics[width=1\linewidth]{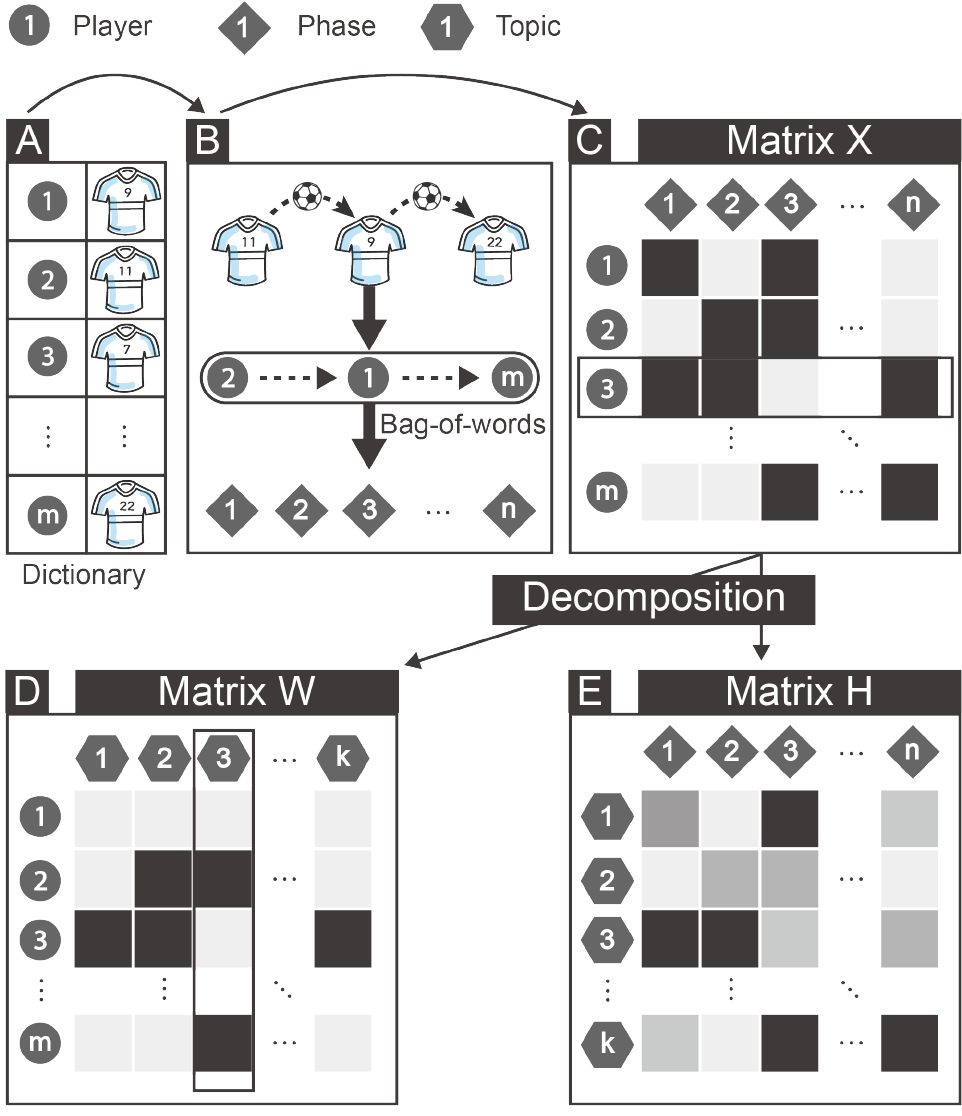}
    \caption{Detecting passing patterns with a topic modeling method. (A) A dictionary of players. (B) A bag-of-words representation of soccer phase. (C) A matrix representing a corpus of players' passing. (D) A matrix representing detected topics (passing patterns). (E) A matrix representing the distribution of passing patterns for each soccer phase.}~\label{model}
    \vspace{-8mm}
\end{figure}

\subsection{Topic-based Pattern Detection}
Given the pattern definition, we aim to decompose a team's players into different groups, in which players tend to pass the ball among one another in a soccer phase. 
We propose a topic-based method for mining passing patterns. We refer to each player as a word, each phase as a document, and each passing pattern as the keyword set of a topic.

\begin{figure*}[htb]
	\centering
	\includegraphics[width=1.0\linewidth]{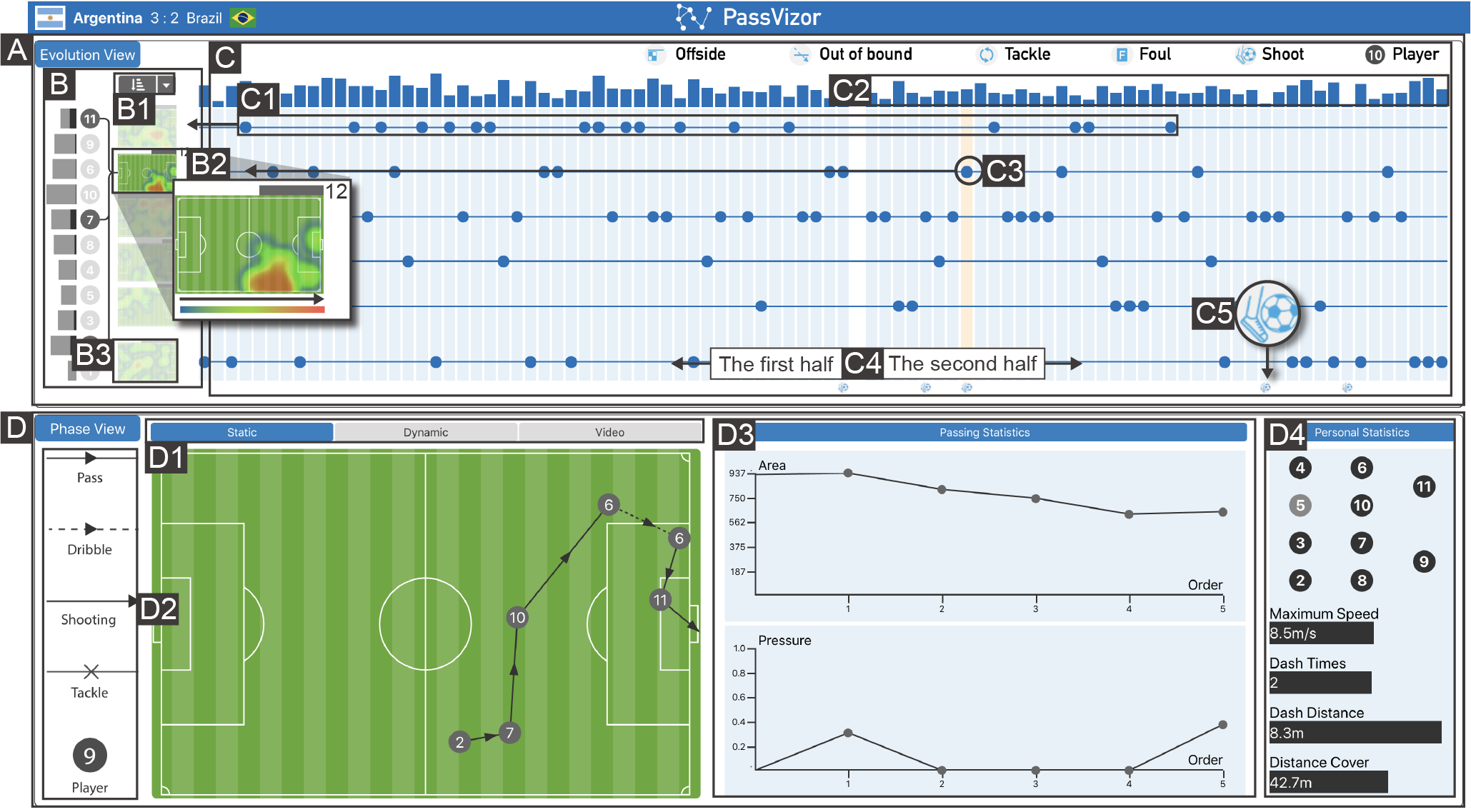}
  \caption{System interface. The system comprises two views, namely, an evolution view (A) and a phase view (D). The evolution view provides a diagram to show a summarization of passing patterns (B) and a  flow (C) to show the temporal distribution of passing patterns over phases. Users can select a phase (C3) and the detailed information, including the passing process (D1) and statistics (D3, D4), can be seen in the phase view (D).}~\label{interface}
  \vspace{-8mm}
\end{figure*}

As shown in Fig.~\ref{model} (A), we first build a dictionary of players.
For each soccer phase, we extract the players involved in each pass and transform a list of passes into a player sequence (Fig.~\ref{model} (B)). This sequence is analog to a sequence of words, and we refer to it as a document.
With the player dictionary, we convert each player sequence into a one-hot vector using the bag-of-words representation (Fig.~\ref{model} (B)).
Here we do not use tf-idf to adjust the weight of players since the meaning of tf-idf is quite confusing in the soccer context.

From multiple soccer phases, we can obtain a list of documents which form a special corpus of passing.
We refer to this corpus as $X \in \mathbb{R}^{m \times n}$, where $n$ is the number of soccer phases and $m$ is the number of players (Fig.~\ref{model} (C)). 
We employ the nonnegative matrix factorization (NMF) \cite{DBLP:conf/nips/LeeS00} to extract the topics because it can easily support parallel accelerations. However, we do not limit the usage of other topic modeling methods. To clarify the detection process, we first provide an illustration of the traditional NMF. The detection can be modeled as:
\begin{flalign}
  \min_{W, H}\|X - W \cdot H\|, \quad s.t.\ W > 0, H > 0
\end{flalign}
where $W \in \mathbb{R}^{m \times k}$, $H \in \mathbb{R}^{k \times n}$, $\|\cdot\|$ is the $L^2$ distance, and $k$ is the number of topics. Topics are represented as the columns of $W$, which are the distribution of players (Fig.~\ref{model} (D)). To transform this distribution to a passing pattern, we can apply the method of extracting keywords (key players) from the topic. As these extracted player groups might share the same players, the overlapping problem is resolved. To determine the passing pattern of each soccer phase, we can utilize matrix $H$. Each column of $H$, $\textbf{c}_i$, encodes the topic distribution for the $i$-th soccer phase. We follow previous methods and assign each soccer phase to the topic with the largest proportion. With this approach, we can obtain multiple passing patterns and the pattern labels for the soccer phases.



\textbf{Preprocess.} We first divide the soccer phases into two categories, namely, the phase with a \textit{counter-attack} and the phase with a \textit{build-up} strategy. Specifically, a \textit{counter-attack} represents a straightforward playing style, i.e., the attack is completed within only a few passes, whereas build-up prefers a control of the ball possession by a series of passes. Due to their different characteristics, the experts aim to separate these two kinds of phases when conducting the tactical analysis. We manually label the soccer phases by inspecting the video.

\textit{Flexible definition of words.}
The definition of a word is flexible. Specifically, users can also apply this topic-based method to investigate the spatial information of passing patterns. For example, we can discretize the spatial position into spatial regions and transform the passing in each phase into a sequence of regions. Referring each spatial region as a word, we can thereby detect passing patterns that characterize the relationship between different spatial regions of the passing. We decide to refer to the player identity as a word since it can provide detailed information of personal characteristics and part of the spatial information, i.e., players' spatial movements are largely affected by their roles in formations, which can be revealed by player identities.

\textit{Substitutions.} 
We can enlarge the player dictionary to handle the substitution. For example, a team of the Premier League could totally have 25 players. We can create a dictionary of 25 words and obtain the Bag-of-Word (BOW) representation of each passing sequence accordingly. Topics from these BOWs contain the weight of substitutes and can be used to explain substituted players’ passing patterns.

\section{Visual Design}

PassVizor is composed of: (a) an \textit{Evolution view} (Fig.~\ref{interface} (A)) for visualizing the topic-based passing patterns and their temporal distributions over soccer phases; (b) a \textit{Phase view} (Fig.~\ref{interface} (D)) for visualizing the fine-grained information of specific phases.
In the \textit{Evolution view}, we provide a \textit{pattern diagram} ((Fig.~\ref{interface} (B))) to visualize the detected passing patterns.
The dynamic change of passing patterns are depicted by a \textit{pattern flow} (Fig.~\ref{interface} (C)), where users can find the temporal distributions of passing patterns over phases.
The required context (e.g. key events and defense) for analyzing passing patterns is shown accompanied by the temporal phases.
By selecting phases in the \textit{pattern flow}, users can examine the detailed process of passing with the \textit{Phase view}, where we provide the positions of actions involved in passing as well as the derived statistics.
Specifically, the attacking direction of the target team is set from left to right and we normalized players' positions according to the direction.
A detailed description of PassVizor is as follows.

\subsection{Evolution View}

\textit{\textbf{Pattern diagram.}} 
In this diagram (Fig.~\ref{interface} (B)), we intend to visualize the characteristics of passing patterns (M1 and M2).
The characteristics are twofold: one refers to the players involved, and the other refers to the spatial context.
As shown in Fig.~\ref{interface} (B), each node at the left represents a specific player and each soccer pitch encodes a passing pattern. 
Users can hover on a pattern and the related players are highlighted  (Fig.~\ref{interface} (B2), Player 11 and 7).
Specifically, we connect a player with a pattern by links if the player is involved in that pattern (Fig.~\ref{interface} (B2)). In particular, the pattern at the bottom (Fig.~\ref{interface} (B3)) encodes the passing of \textit{counter-attack} while the others encode the \textit{build-up}.  We decided to use a node-link diagram to encode the detected patterns since it is more intuitive compared with the matrix (i.e., a common representation for topics) when the number of topics is not very big.

We use a heatmap (Fig.~\ref{interface} (B2)) to encode the spatial information of passing.
The spatial information of passing can be represented by a trajectory of players' positions.
Sacha et.al. \cite{Sacha2017cgf} regard the start and the end position as the most important information of a trajectory. 
We follow this simplification and collect the start and end positions of all passing trajectories for a passing pattern.
We then visualize these positions on the soccer pitch with a heatmap (Fig.~\ref{interface} (B2)).

We also show additional statistics of passing patterns in this diagram. 
The bar next to the player (Fig.~\ref{interface} (B)) encodes each player's number of passes.
When hovering on a pattern, a dark bar (Fig.~\ref{interface} (B)) is presented to show the number of a player's passes in that passing pattern.
This can help users confirm the importance of a player.
The bar on each pattern (Fig.~\ref{interface} (B2)) encodes its frequency in a match.

\textit{\textbf{Pattern flow.}}
We employ a timeline-based visualization to show the temporal distribution of passing patterns (P1-P3). Each circle (Fig.~\ref{interface} (C3)) represents a phase and we place them from left to right in a chronological order. The vertical position of a circle is aligned with the corresponding passing pattern in that phase (Fig.~\ref{interface} (C1)).
Phases of the first half and the second half are separated (Fig.~\ref{interface} (C4)).
Users can hover on each circle and the corresponding passing pattern will be highlighted. 
This can help users identify the change of passing patterns.

To provide the context of each phase (P2), we place a bar on the top to show the defense in each phase (Fig.~\ref{interface} (C2), the higher the height the worse the defense).
We use the covered region of the opponents to represent the defense.
A small covered region means that a team is using a good defense formation.
We also place event glyphs (Fig.~\ref{interface} (C5)) at the bottom to show the ending (P3).
We only show the event of shooting at the beginning of the analysis to reduce the cognitive load.
We use the metaphor to design the game event glyphs. For example, we design the kicking behavior to encode the shooting event and use the icon of conversion to encode the substitutions. We also borrow existing designs from common soccer icons (e.g., cards and goals).

\begin{figure}
	\centering
	\includegraphics[width=1\linewidth]{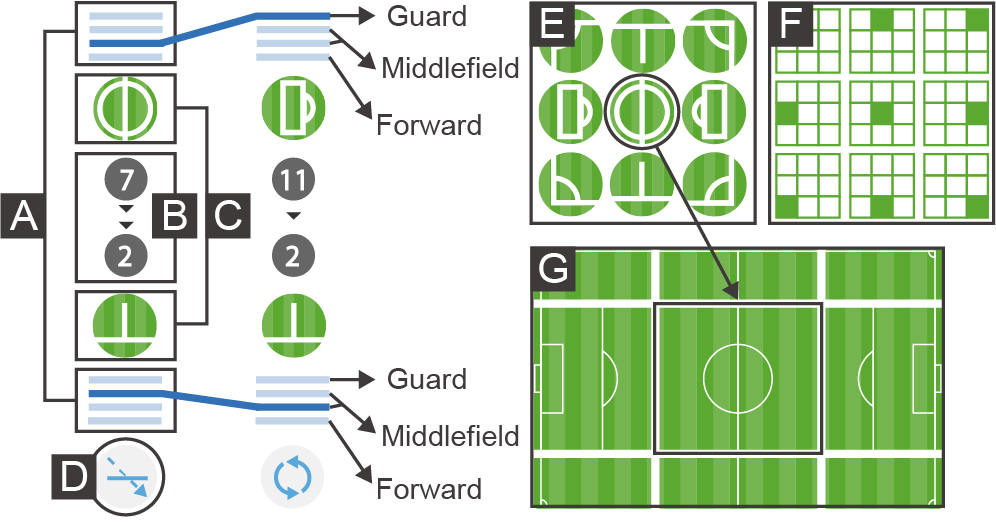}
	\caption{Design for visualizing the passing of a phase. (A) Formations. (B) The first passer (upper one) and the last receiver (lower one). (C) The spatial region. (D) The end event. (E) The spatial glyphs for all regions. (F) An alternative of spatial glyphs.}~\label{alternative}
	\vspace{-8mm}
\end{figure}

According to the change of passing, users could be interested in certain passing patterns.
For an in-depth analysis of a passing pattern, we allow users to zoom in by clicking the pattern (Fig.~\ref{interface} (B2)) and a fine-grained visualization of the corresponding phases will be shown (Fig.~\ref{case1_2}).
The temporal distribution of the selected pattern is preserved on the top (Fig.~\ref{case1_2} (A)).
In this visualization (Fig.~\ref{alternative}), each column, which is composed of multiple glyphs, provides a multivariate summarization of the passing in each phase.
There are a set of techniques that can visualize multivariate data, such as parallel coordinates and projections.
However, the parallel coordinates can not preserve the spatial information and the result of projection is hard to interpret.
Hence, we decide to use glyph techniques which have high efficiency in encoding the multivariate information \cite{DBLP:journals/ivs/ChungLPBGLC15} and good readability for users.

\textbf{Summarization of a phase}. According to experts' suggestions and previous works \cite{Sacha2017cgf}, the first (the first passer) and the last point (the last receiver) are considered as the most prominent information of a phase.
Therefore, we visualize the multivariate information of the two points respectively.
For a phase, we provide the formation (Fig.~\ref{alternative} (A)),  the identity (Fig.~\ref{alternative} (B)), and the position (Fig.~\ref{alternative} (C)) of the two points respectively. The end event (Fig.~\ref{alternative} (D)) is placed at the bottom. The density of triangles in the middle encodes the number of passes in this phase. By juxtaposing the multivariate information of multiple phases, users can obtain a deeper understanding of passing patterns.

\textit{Formation.} 
The formation glyph (Fig.~\ref{alternative}(A)) shows the numbers of formation lines.
For example, 4-4-2 is encoded by three lines while 4-2-3-1 is encoded by four lines.
The highlighted line (Fig.~\ref{alternative} (A)) indicates the position of the first/last in the formation. This design is derived from previous work \cite{forvizor} who used a Sankey to encode the formation. 
We drop the thickness channel of lines (which is used to encode the proportion of players in each line) for two reasons. First, considering the limited visual space, it is hard for users to perceive the thickness of lines. Second, the target of the formation glyph is to notify users which kinds of players are possessing the ball (e.g., the strikers or the midfielders). Such information is encoded by the position. For these two reasons, we propose the current design.

\textit{Spatial position.} The shape of the glyph (Fig.~\ref{alternative} (E)) indicates the spatial region in which the pass is conducted.
Specifically, we divide the soccer pitch into nine different spatial regions (Fig.~\ref{alternative}(G)) and design glyphs to represent the regions respectively.
The design of the glyph is inspired by the most prominent visual features in each spatial region.
For example, we use the circular shape in the middle of the pitch to represent the midfield (Fig.~\ref{alternative} (E)). 
A mapping between the glyph and the corresponding spatial region is available in Fig.~\ref{alternative} (E).
We have designed two design alternatives to encode the spatial regions.
The first one is the most primitive idea, using nine panels to encode the spatial regions respectively (Fig.~\ref{alternative} (F)).
For example, to represent the midfield, we highlight the panel at the center.
However, given a relatively large number of glyphs, it would be difficult for users to discern them as they would look similar in small, let alone distinguish from each other.
To improve the discrimination, we tried to embed the distinctive feature of each spatial region into the design of the glyphs and therefore propose the current design (Fig.~\ref{alternative} (E)). According to the expert comment, it is easy for them to understand the encoding of this glyph design since they are familiar with the embedded glyph shape.

\textit{Interactions.} This view supports interactions as follows.
\begin{compactitem}
	\item \textbf{Ranking.} Users can rank the passing patterns by clicking on the ranking button (Fig.~\ref{interface} (B1)). The layout of the pattern flow will be re-organized, which can help users more easily perceive the temporal patterns. Users can select different ranking attributes, such as the number of occurrences and the number of shootings.


	\item \textbf{Hovering.} Users can hover on the player node (Fig.~\ref{interface} (B)) and the soccer phase that contains this player will be highlighted (in both the two levels of pattern flow). This can help users analyze the passing characteristics of a player.
	
	\item \textbf{Switching.}  We allow users to switch between the passing of the two teams in a match. Users can click on the team name (Fig.~\ref{interface}) to inspect the passing and the dynamic changing of the passing pattern of the two teams respectively.

\end{compactitem}

\subsection{Phase View}

This part is provided to present the soccer phase selected in the evolution view in detail.
We provide three different modes to facilitate the passing analysis, namely, the static mode (Fig.~\ref{interface} (D1)) to show a static summarization of all passes, the dynamic mode (Fig.~\ref{interface} (D1)) to visualize the animated movements of the players during passing, and the video mode (Fig.~\ref{interface} (D1)) to present the raw soccer video. The time interval of the three modes is coordinated.
As shown in Fig.~\ref{interface} (D1), in the static mode, we use a node-link diagram to visualize the whole process of players' passes in a soccer phase.
Players are placed according to their position when they pass or receive the ball.
We use the solid line to encode a pass between the two connected players (Fig.~\ref{interface} (D2)) and a dashed line to encode the movement of a player when dribbling (Fig.~\ref{interface} (D2)).
To reduce the visual clutter, we remove the slight movements of players.
For the end event of the phase, we use an arrow to show the shooting event, a cross to show the intercepted pass.
Users can also select a specific pass to see the context information at that time.
When users click on Player 10, the positions of 11 opposing players (when Player 10 passes the ball) are shown.
With the information, users can quickly learn how Player 10 is being defended.

For a deep analysis, users require quantitative context information to know the actual situation that faced by players during a passing.
Therefore, we provide a statistical table and a set of coordinated interactions to show the necessary context. In the statistical table (Fig.~\ref{interface} (D3)), we provide the following indicators to facilitate the analysis.
\begin{compactitem}



	\item \textbf{Covered area.} This describes the covered area of the opposing players. We calculate this based on the polygon generated by players' positions on the pitch. Generally, a smaller value of covered area represents a better defense.

	\item \textbf{Pressure value.} This is to estimate the pressure imposed on the ball-possessing player. A closer defender would pose a higher pressure on the player. We calculate this value based on Andrienko et al.'s method \cite{Andrienko2017pressure}.
\end{compactitem}

Apart from the statistics of passing, we further provide statistics (e.g., maximum speed, dash distances, etc) of individual players (Fig.~\ref{interface} (D4)). Players are placed according to the lineup of the team. Users can click on the player to show the corresponding statistics. 

\section{Expert Interviews and Discussion}

\begin{figure}[t]
	\centering
	\includegraphics[width=1\linewidth]{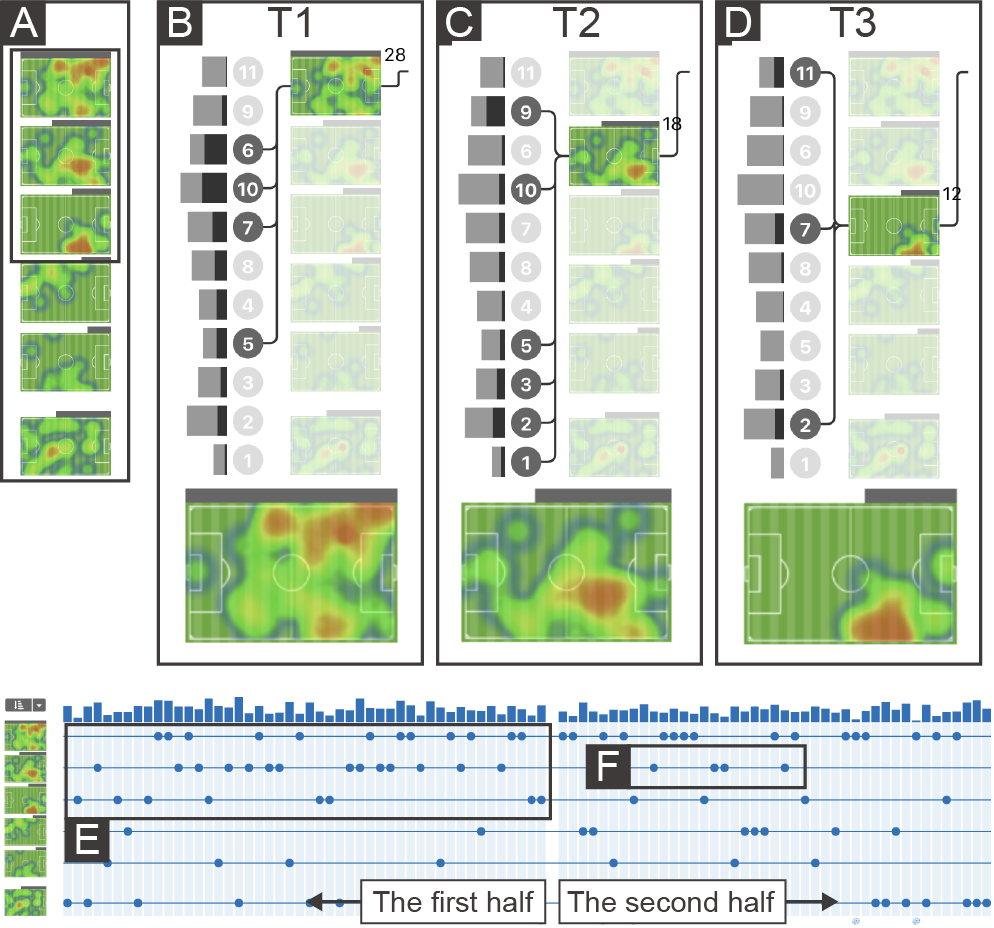}
	\caption{The change of Argentina's passing. (A) Three frequent passing patterns (B), (C), and (D). (E) and (F) show the temporal distributions of the three passing patterns.}~\label{case1}
	\vspace{-8mm}
\end{figure}

\begin{figure*}[t]
	\centering
	\includegraphics[width=1\linewidth]{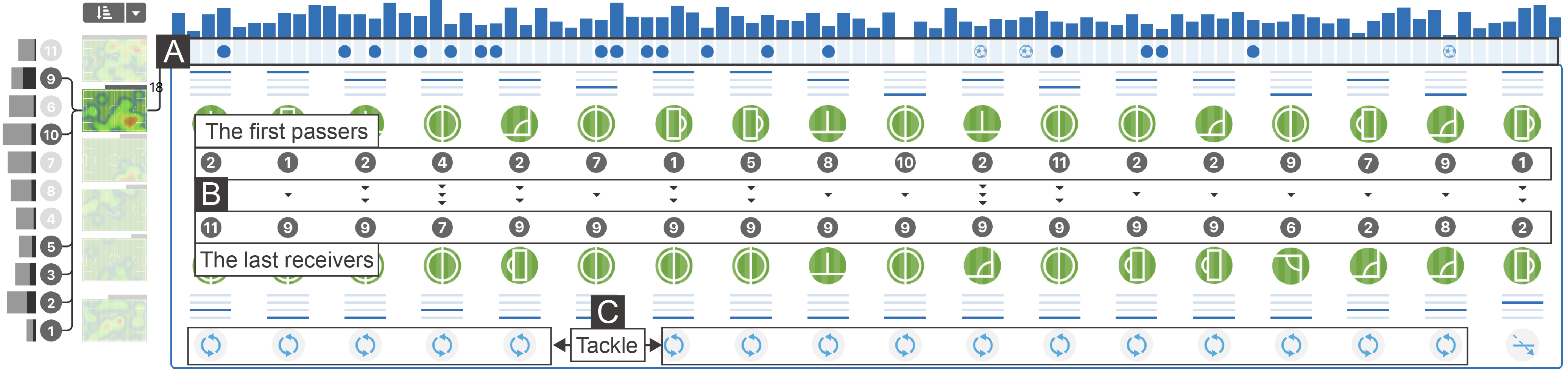}
	\caption{The soccer phases of a passing pattern. (A) The temporal distribution of selected passing patterns over soccer phases. (B) The first passer and the last receiver in each soccer phase.  (C) The ending event of each soccer phase.}
	\label{case1_2}
	\vspace{-5mm}
\end{figure*}

In this section, we demonstrate the usability of PassVizor with expert interviews.
We invited two experts to conduct an in-depth analysis of soccer passing behaviors.
The first expert (referred to as E1) was a senior sports analyst.
The second expert (referred to as E2) was a senior coach with a coaching certificate issued by a continental football confederation.
For the evaluation, we used two matches (Argentina versus Brazil and Argentina versus Peru) of the Under-15 Football Championship.
We invited the experts to analyze the passing of Argentina.
The experts knew about the match result (e.g., the score, the winner, and the type of goals) before the analysis. Nevertheless, the passing data was totally new for them. It was also the first time for them to analyze a match from the perspective of passing.

During the interviews, we first provided an introduction to the usage of PassVizor.
Thereafter, the experts were asked to conduct case studies with PassVizor.
After case studies, we collected the experts' feedback and suggestions on usability.
From the case studies, the experts derived several insights into the change of passing patterns.
A summarization of the insights and the experts' comments is as follows.

\subsection{Expert Interviews}


\subsubsection{Change of Passing for Improving Attack Efficiency}
This insight was obtained from the match Argentina versus Brazil.
The experts inspected the pattern diagram first to see how many patterns are found. 
They noticed six passing patterns (Fig.~\ref{case1} (A)), and one of the patterns, the counter-attack, is separated from others (M1).
They appreciated this design and commented ``\textit{this could help me easily distinguish the two types of passing}''. 
The experts decided to learn the passing patterns of build-up first.
They ranked the passing patterns based on the frequency to find the frequent passing patterns.
As shown in Fig.~\ref{case1} (A), from the bar on each pattern, the experts identified three main passing patterns.
The experts then found that the top three patterns (Fig.~\ref{case1} (A)) have some high-density areas on the heatmap (M2).
They commented \textit{``It seems a set of players insist on attacking the same area during the match.''} and regarded these three as important patterns.
The experts then hover on the three patterns to learn more details.
Based on the links and the bars, the experts deduced that the three passing patterns were 1) a tactic for attacking midfielders 6 and 10 (Fig.~\ref{case1} (B), forwards were rarely involved according to the darker bar, referred as T1), 2) a tactic for the forward 9 (Fig.~\ref{case1} (C), referred as T2), and 3) a tactic for the other forward 11 (Fig.~\ref{case1} (D), referred as T3), respectively.


Specifically, T1 highly emphasized the involvement of the side midfielder 6 and the center midfield 10.
Moreover, based on the heatmap, the experts learned that the tactic for attacking midfielders was frequently used in the left flank (shown in Fig.~\ref{case1} (B), which is consistent with the player identities) and the tactics for the two forwards were conducted in the right side of the pitch (Fig.~\ref{case1} (C), (D)).
Combining their domain knowledge, the experts deduced that the three passing patterns comprised of the main attacks from Argentina.

After that, the experts were interested in the usage of these passing patterns and switched to the pattern flow for further explorations (P1). 
From the distribution of circles, the experts found that the three patterns contribute to most of the phases in the first half (Fig.~\ref{case1} (E)).
However, there was a significant decrease in the usage of T2 in the second half (Fig.~\ref{case1} (F)).
To investigate this decrease, the experts clicked on T2 to inspect the passing in detail.
From the summarization of each phase (Fig.~\ref{case1_2} (B)), the experts learned that the first passers were diverse, including multiple guards, while more than half of the last receivers were forward 9. 
This helps the experts obtain a deeper understanding of T2.
Combining this information with the heatmap of T2, they commented \textit{``T2 may be a strategy of passing the ball to forward 2 on the right side for the attack''}. 
However, according to the event glyphs (Fig.~\ref{case1_2} (C)), the experts found that T2 was inefficient as most of the passes were tackled by the opponents and none of them created a shooting chance. ``\textit{Argentina may realize that this kind of passing was not effective against the defense of Brazil in the half-time interval and decided to reduce this kind of attacks in the second half}'' the experts commented.
In this process, the experts learned the transfer of the main passing patterns by utilizing the pattern diagram and the pattern flow.


\begin{figure}[!b]
	\centering
	\vspace{-4mm}
	\includegraphics[width=1\linewidth]{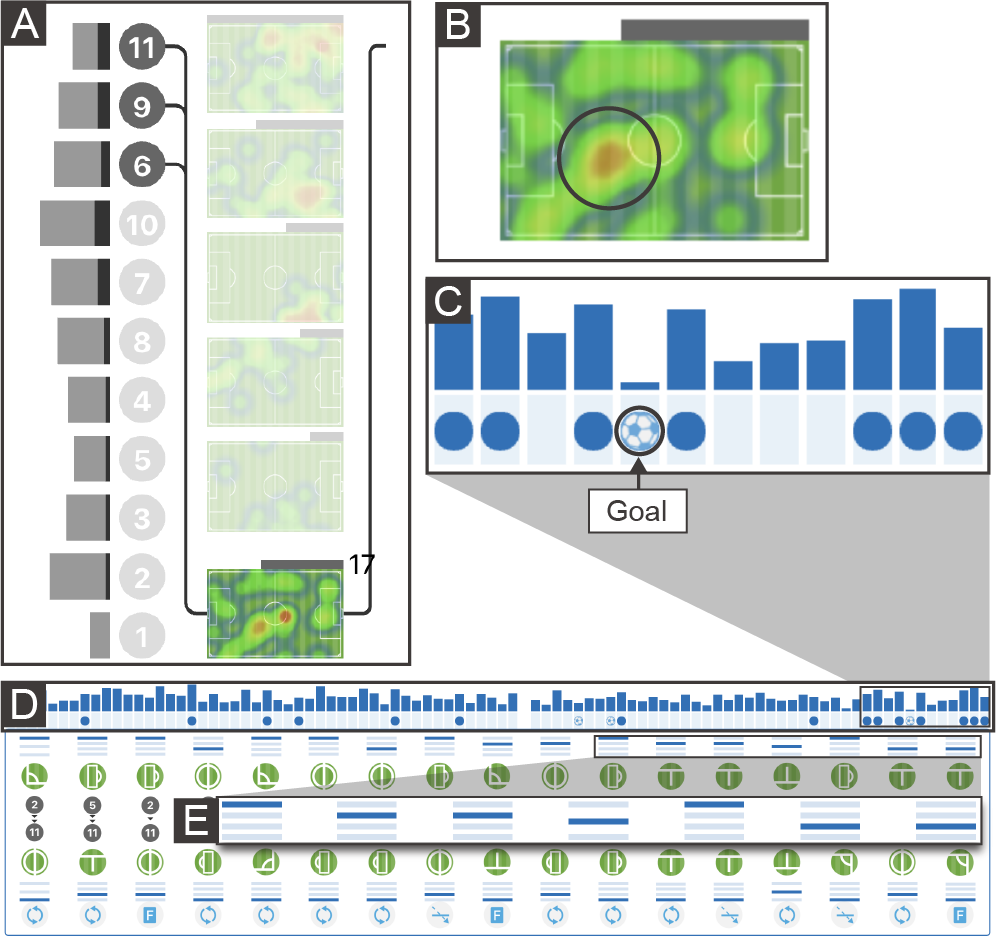}
	\caption{The counter-attack of Argentina. (A) The frequent players in the counter-attack. (B) The spatial information of counter-attack. (C) The frequent employment of counter-attack near the end of the match. (D) The faced defense of counter-attack. (E) The formation line of the first passer of counter-attack.}~\label{case2}
	\vspace{-3mm}
\end{figure}

\subsubsection{Counter-Attack for Holding The Lead}
This insight was obtained from the match Argentina versus Brazil.
E2 was especially interested in the usage of the counter-attack.  
E2 first hovered on the corresponding soccer pitch and found the distribution of players in the passing of counter-attack. (M2) Specifically, E2 discovered three important players, including two forwards (Player 11 and 9) and a midfielder (Player 6) (Fig.~\ref{case2} (A)). 
According to the player identities, E2 deduced that most of the counter-attacks were accomplished by the three players. 
E2 then inspected the heatmap to learn the position of the counter-attack. E2 stated that generally, the counter-attack would be caused by a recovery of the ball possession, i.e., when players successfully intercept the ball during defense, they will pass forward (to midfielders or forwards directly) in a high speed to attack. Hence, the beginning (where they intercept) and the end position (where they try to attack) are significant for learning and evaluating the counter-attacks. When inspecting the heatmap, E2 was attracted by a highlighted point near the midfield (Fig.~\ref{case2} (B)). \textit{``It seems that most counter-attacks were launched from the backfield''} E2 commented.

E2 turned to the pattern flow to see when the counter-attack was applied and the corresponding results (P1). 
As shown by the defense bar (Fig.~\ref{case2} (D)), most counter-attacks were conducted when the covered areas of the opponent were comparatively high.
This represented that the decision to employ counter-attack in these soccer phases as appropriate (P2).
E2 further discovered that the counter-attack was occasionally used in most of the time when examining the temporal distribution (Fig.~\ref{case2} (D)).
However, near the end of the second half, E2 found a list of consecutive circles (Fig.~\ref{case2} (C)), representing that the counter-attack was frequently used.
According to the match events, E2 learned that such transition to the counter-attack began before Argentina obtaining a one-goal lead (Fig.~\ref{case2} (C)).
Moreover, after Argentina led the game, they persisted on employing the counter-attack in the final stage of the match. E2 commented \textit{``the team decreased the times of ball possessions to reduce the probability of making mistakes and focused on defending against the opponents in the final stage''. }(P3) For more details of the counter-attack, the experts further clicked on the soccer pitch and zoom into the detailed level of the pattern flow.

Based on previous observations, E2 focused on analyzing the counter-attack in the final stage. 
Specifically, in the last three soccer phases, the experts found that the formation line of the first passer was moving forward (Fig.~\ref{case2} (E)), showing a shift of the position of interception.
This represented that more players were involved in the defense.
From the match events, E2 further noticed that Argentina conducted a counter-attack before scoring the last goal. From the glyph of this counter-attack soccer phase, E2 found that this attack was initiated by Player 4 in the right flank and ended by the Player 6 with an out-of-bound (I1). To learn the detail process of this attack, E2 further inspect this phase in the phase view. E2 found that although this attack was stopped by the opponent, it created a corner kick for Argentina, which was the cause of the last goal. \textit{``Although the change to counter-attack did not create shooting chances directly, it still made significant contributions to the win of Argentina''} E2 commented.
In this process, the experts learned the change to counter-attack by utilizing the pattern flow and the phase view.

\begin{figure*}[t]
	\centering
	\includegraphics[width=1\linewidth]{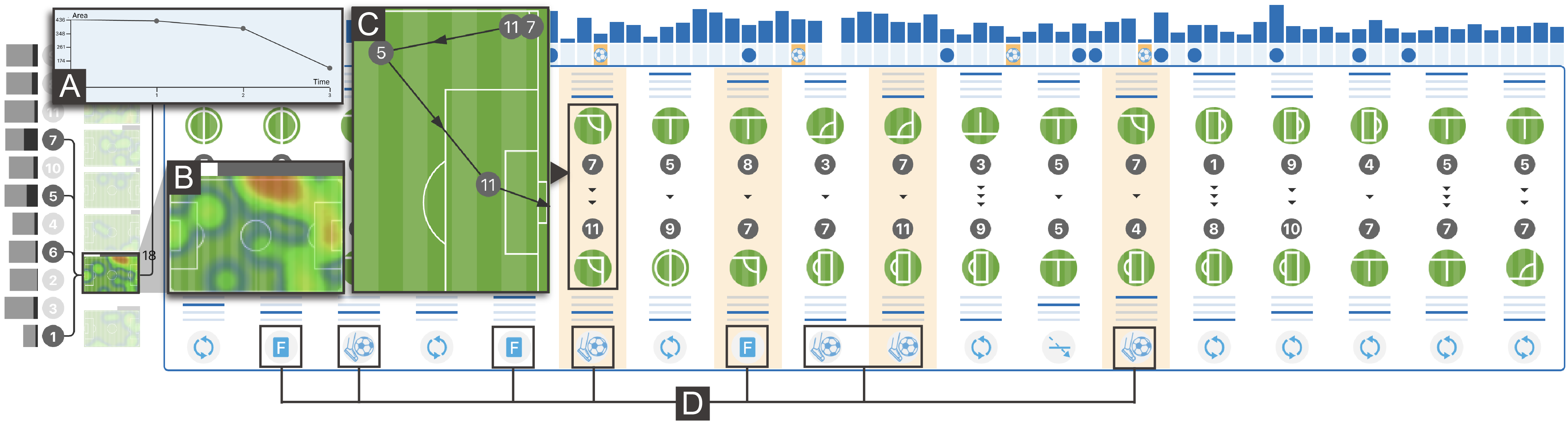}
  \caption{The efficient passing pattern of Argentina. Hovering on the soccer pitch, the experts learned the spatial information and involved players of this passing pattern (B). A list of soccer phases with shooting events was also highlighted, representing that this pattern was efficient for attacks (D). By inspecting a soccer phase, the experts knew that Argentina conducts these efficient attacks by set-piece (C), in which Player 7 was the corner kicker. From the passing statistics (A), the experts found a core player (Player 5), which conducted a pass that penetrated the defense (caused a sudden drop of the defense area of the opponent) and led to a goal. }
  ~\label{case3}
	\vspace{-7mm}
\end{figure*}

\subsubsection{Set-piece as The Most Effective Attack}
This insight was obtained from the match Argentina versus Peru.
When analyzing Argentina's passing in this match, the experts were immediately attracted by the large numbers of shooting in the pattern flow.
This showed that Argentina had posed great pressure on Peru and controlled the match.
To learn which passing pattern was efficient for attacks (i.e., create shooting chances), the experts hovered on the pitch of each passing pattern and found that a passing pattern that mainly comprised of Player 7 and Player 5 created most shooting chances for Argentina (Fig.~\ref{case3} (D), M2).
According to the heatmap, the main active region was in the left flank (Fig.~\ref{case3} (B)). \textit{``This was consistent with the player identities as Player 7 was the left side midfielder and Player 5 was the left full back''} the experts commented. The experts then turned to the pattern flow to further analyze this passing pattern (P1). By inspecting the events (Fig.~\ref{case3} (D)), the experts found that this tactic not only created shooting chances but also caused several fouls, which further confirmed its efficiency for attacks.
Specifically, hovering on these successful soccer phases with shooting and fouls, the experts learned that all of the goals of Argentina were scored by this pattern (highlighted phases in Fig.~\ref{case3}).
The experts then inspected these phases with goals in detail to learn the passing.
From viewing the information provided by the three modes in phase view, the experts discovered that most goals were created by a set-piece that involved a similar set of players in which Player 7 was the core player. Player 7 was the corner kick taker for Argentina (Fig.~\ref{case3} (C)), which made a successful pass to the teammates (I2). Moreover, instead of conducting general corner kicks, the experts also found that Argentina performed a short corner and scored a goal (Fig.~\ref{case3} (C)). According to the statistics (Fig.~\ref{case3} (A)), the experts found that the key pass was performed by Player 5 as this pass penetrated the defense of Peru (A sudden drop of the covered area). In this process, the experts learned the efficient passing pattern and its core players by utilizing the pattern diagram, the pattern flow, and the phase view.

\subsubsection{Expert feedback}
After the case studies, we interviewed the experts and collected their useful feedback and suggestions on our system. 
Overall, the experts were satisfied with PassVizor.
The experts felt that the system was useful, as it could support an in-depth analysis of a team's passing. Compared with traditional video analysis, this system can help them significantly reduce the time for learning the passing patterns. 
Specifically, the experts appreciated the phase-level presentation of the soccer passing. 
They commented that each phase could be regarded as a unit of attacks.
They would like to analyze the passing phase by phase since they do not want to miss important details and variations.
But the phase-level analysis with videos is too time-consuming, which forces them to use aggregation analysis.
The transition to the details of soccer phases in the pattern flow was also smooth and reasonable.
Nevertheless, with this tool, the experts now can inspect multiple soccer phases in a short time, which can be a great complement to their current analysis workflows. 
The experts also appreciated the spatial context provided by the spatial glyph as traditionally, they could only obtain this information by inspecting the raw video. 

\textbf{Suggestion.} E2 suggested that the color theme of the system can be changed according to the target team of analysis (e.g., according to the color of the team logo) as they were much more familiar with such color encoding.
This can also help analysts more easily distinguish between the switch of the targeted team. E1 commented that the pattern flow could be improved by providing flexible filtering. Although the current design is suitable for summarizing the passing, they hope to have a set of interactions that can help them filter soccer phases and compare the passing of these phases in detail. For example, they want to compare the passing of different phases which have shooting chances.

\subsection{Discussion}
In this work, we focus on analyzing the dynamics of passing patterns to reveal the adjustment of passing (including self-adjustments by players and proactive adjustments by coaches). We contribute a topic-based passing modeling approach to capture the passing pattern and propose a glyph-based visualization to demonstrate the multi-variate context of passing in multiple soccer phases.
With PassVizor, users are enabled to identify the valuable changes of passing that contribute to the lead and examine the detailed passing process to disclose the key pass and core players.
The expert interviews demonstrate the usability of our system.

\textbf{Lessons Learned.} One applicable lesson is the representation of soccer phases. Initially, we represent each soccer phase as a network of players and regard the dynamics of passing patterns as a temporal passing network. We try to use a temporal network embedding method to address the problem. However, different from the continuously evolving process of general temporal networks, the change of the network of each soccer phase is discrete as one soccer phase is not built on the previous phase. The other lesson is the visualization of the spatial region in the sports domain. 
Although our spatial glyph is mainly designed for the soccer data, the strategy of utilizing the intrinsic signs on courts can be applied for other sports.
In many kinds of sports, the court is also divided into different regions with different signs according to the disciplines.
For example, in basketball, the court is divided into different areas with different shapes, such as the free-throw lane and the area beyond the three-point line.
Integrating such visual features can improve the usability and the comprehensibility of visualizations (according to the expert comments), thereby facilitating the analysis.

\textbf{Scalability.} 
Although the current system is designed to analyze the passing of one match, both the passing pattern detection and the passing visualization are able to be adapted to the analysis of multiple matches. The adaptation mainly depends on two issues, i.e., the detection of multiple matches’ passing patterns and the visualization of soccer phases in multiple matches.
For the passing pattern detection, multiple matches of a team share the same player dictionary. Hence, it is possible to conduct topic-based passing pattern detection on multiple matches’ passes and find a shared set of passing patterns. Moreover, topic modeling has been widely applied to process large text datasets. Hence, the model can even be used to analyze a season’s matches. If there are too many patterns, we can control the parameter of topic models to acquire a limited number of the most significant passing patterns. 
Moreover, since topic modeling has been widely applied to process large text datasets, it is possible to apply the topic-based passing pattern detection to one or more seasons' matches.
For the visualization of soccer phases, there will be multiple lines of pattern flows when analyzing multiple matches. Regarding passing patterns as the vertical axis and the order of soccer phases as the horizontal axis, we can transform a pattern flow to a line chart and investigate the different evolution by comparing the trend of lines. To ease the comparison, we can utilize sequence alignments to automatically find similar passing trends between multiple matches and highlight these similar time slots in the line chart. Hence, it is possible to adapt the system to multiple matches.

\textbf{Reproducibility.} This work is the same as many topic modeling works in terms of reproducibility. The major parameter here is the number of topics. In our cases, since the data is not very large, we tried different numbers of topics and inspected the result to choose the best parameter. Specifically, as we state in the last paragraph of Section 5, whether a soccer phase is a counter-attack or build-up is manually labeled. We then input all the phases of build-up into the topic model. Currently, as the data is not big, we can obtain the result of topic modeling in only a few seconds and it was easy for us to find that an appropriate number of topics. For a large dataset, there are several widely adopted methods for automatically choosing the parameter. For example, when using LDA, HDP (Hierarchical Dirichlet Processes) \cite{hdp} can be used to determine the number of topics. There are also heuristic approaches\cite{zhao2015heuristic} to set the parameter. Therefore, our work can be reproduced and we will use a larger dataset to illustrate the reproducibility in the future.


\textbf{Limitations.}
The limitations are twofold.
The first limitation lies in the limited consideration of the opponent team.
The experts suggested that the analysis can be further improved by considering the behaviors of the opponents, which can disclose more detailed features of the passing.
For this issue, we regard the passing patterns as the observed result that is influenced by different factors, such as opponents’ defense strategy and players’ personal status (e.g., fatigue). To consider these factors, we can model the relationship between players’ passes and these factors. For example, for each soccer phase, we can build the connection between the opponent’s defense strategy and the observed passing pattern. We can learn the probability of using passing pattern A when the opponent applies certain defense strategy. This can answer questions like how the player will pass the ball when the opponent adopts a deep defense.
In the future, we plan to develop appropriate methods for modeling the defense and investigate how to establish the relationship between the defense and the pass.
The second limitation lies in the presentation of the spatial regions of passing behaviors.
Although we have designed intuitive glyphs to encode the spatial regions in soccer, the problem of demonstrating the transition of spatial regions for passing remains challenging.
To create scoring chances, players would progressively move from the backward positions to the forward positions, thereby causing a smooth transition of the spatial regions.
Specifically, different attacking strategies would lead to different transition patterns of the spatial regions.
Thus, it is worth to reveal such transitions for further analysis of passing.

\section{conclusion}
This study characterizes the problem of analyzing the dynamics of soccer passing behaviors.
To support users in recognizing and interpreting the passing dynamics, we propose a topic-based approach to model the change of player identities involved in the passing behaviors of different soccer phases.
Based on the model, we design and implement an interactive visual analytics system called PassVizor to assist users in analyzing the dynamic passing behaviors with sufficient context information.
For the investigation and the comparison of multiple soccer phases, we create a glyph-based design to visualize the multi-variate information of soccer phases, including the player identities, the associated formation, and the corresponding spatial context.
We used expert interviews to clarify and evaluate the usability of PassVizor. In addition to soccer, our method can also be applied for analysis in other sports which contain behaviors similar to passing.

In the future, we plan to integrate the actions and the influence of the opponents into the passing modeling method to more accurately model the dynamic-changing process of passing behaviors. Moreover, we also plan to extend PassVizor to support the analysis of the co-evolving passing behaviors of the two teams in a match as well as the analysis of passing behaviors across multiple matches.

\acknowledgments{
The work was supported by National Key R\&D Program of China (2018YFB1004300 ), NSFC (61761136020), NSFC-Zhejiang Joint Fund for the Integration of Industrialization and Informatization (U1609217), Zhejiang Provincial Natural Science Foundation (LR18F020001) and the 100 Talents Program of Zhejiang University. }

\end{spacing}

\bibliographystyle{abbrv-doi}

\bibliography{references}
\end{document}